\documentclass[runningheads]{llncs}
\usepackage{amsmath}
\usepackage{graphicx}
\usepackage{subfig}
\usepackage{authblk}
\usepackage{graphicx}
\begin{document}

\title{Technical Report of 1:10 Scale Autonomous Vehicle Robot }
%
%

\author{Amirhossein Kheiri Holighi\inst{1} \and
Seyed Sobhan Hosseini Hajibekandeh \inst{1} \and
Amirhossein Gholizadeh Behbahani\inst{1} \and 
Kian Khatibi\inst{2} \and
Saina Najafi Shabestari \inst{2} \and
Ghazal Ghoreishi \inst{2} \and
Aria Dadnavi \inst{2} \and
Saba Sadeghi\inst{2} \and 
Shahriar Karimi Makhsous\inst{2} \and
Matin Jamshidi\inst{2} \and
Mandana Shabanzadeh Nasrolah Abadi\inst{2} \and
Mohammad Hossein Moaiyeri \inst{3}
}

\authorrunning{A. Kheiri, S. S. Hosseini, A. Gholizadeh, et al.}
%
\institute{Faculty of Electrical Engineering, Shahid Beheshti University, Tehran, Iran \\
\email{h\_moaiyeri@sbu.ac.ir}\\
\email{\{a.kheiriholighi, s.hoseinihajibekande, am.gholizadeh, k.khatibi, s.najafishabestari, g.ghoreishi, a.dadnavi, sab.sadeghi, s.karimimakhsous, mat.jamshidi, m.shabanzadeh\}@mail.sbu.ac.ir}\\
\url{https://sbu-team.github.io/}}
\maketitle              

\begin{abstract}
This paper presents Auriga Robotics’ autonomous vehicle, developed at Shahid Beheshti University’s Robotics and Intelligent Automation Lab, as part of the team’s entry for the 2024 RoboCup IranOpen competition. The vehicle is a 1:10 scale car equipped with a custom-designed chassis, a stepper motor for precision, and a range of sensors for autonomous navigation. Key hardware includes ESP32 microcontrollers that manage motor control and sensor data acquisition. The software system integrates computer vision, including YOLOv8 for sign detection and PiNet for lane detection, combined with control algorithms such as the Stanley, PID, and Pure Pursuit controllers. The vehicle's design emphasizes real-time decision-making, environmental mapping, and efficient localization, ensuring its ability to navigate complex driving scenarios.

\end{abstract}
\begin{figure}
\centering
\includegraphics[width=8cm]{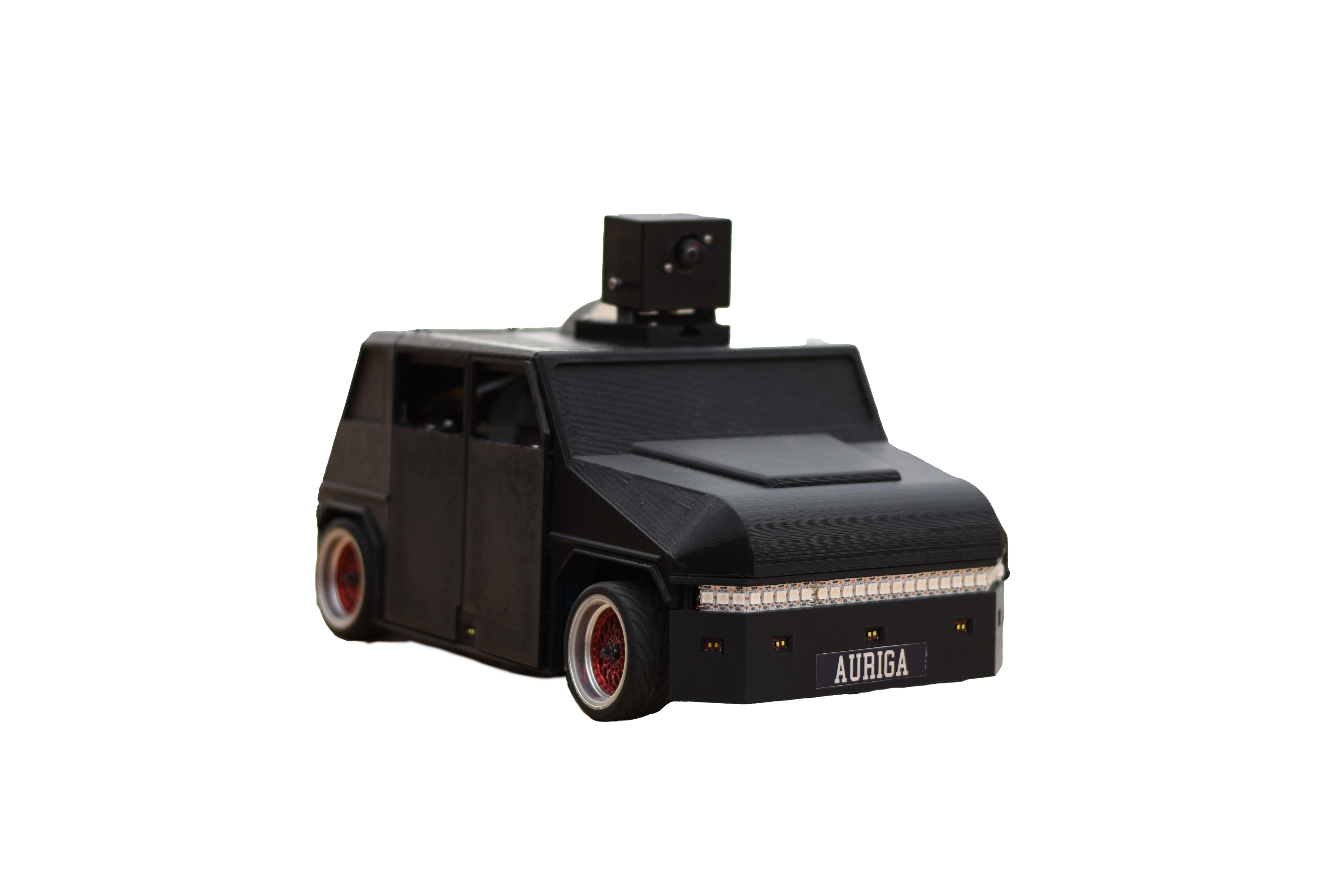}
\caption{Photo of the robot.} \label{fig1}
\end{figure}

\section{Introduction}
Autonomous vehicles (AVs) are at the forefront of innovation in transportation technology, providing solutions for navigating complex environments without human intervention. The development of the Auriga autonomous vehicle focuses on integrating precision hardware with advanced control systems to enable reliable and efficient autonomous driving. The vehicle's design includes a reinforced chassis and a stepper motor-based drivetrain that supports enhanced maneuverability and power transmission. Equipped with an array of sensors, including cameras and range sensors, the vehicle achieves real-time environmental awareness. Combined with a robust software stack, featuring cutting-edge computer vision algorithms and control systems, this AV demonstrates significant capabilities in navigating dynamically changing environments with accuracy and responsiveness. The following sections explore the detailed hardware, software, and control strategies that enable the vehicle to perform complex navigation tasks autonomously.

\section{Hardware Description}
\subsection{Mechanical Design}
\subsubsection{Car Chassis}
\begin{figure}
    \centering
    \subfloat[\centering]{{ \includegraphics[width=5.5cm]{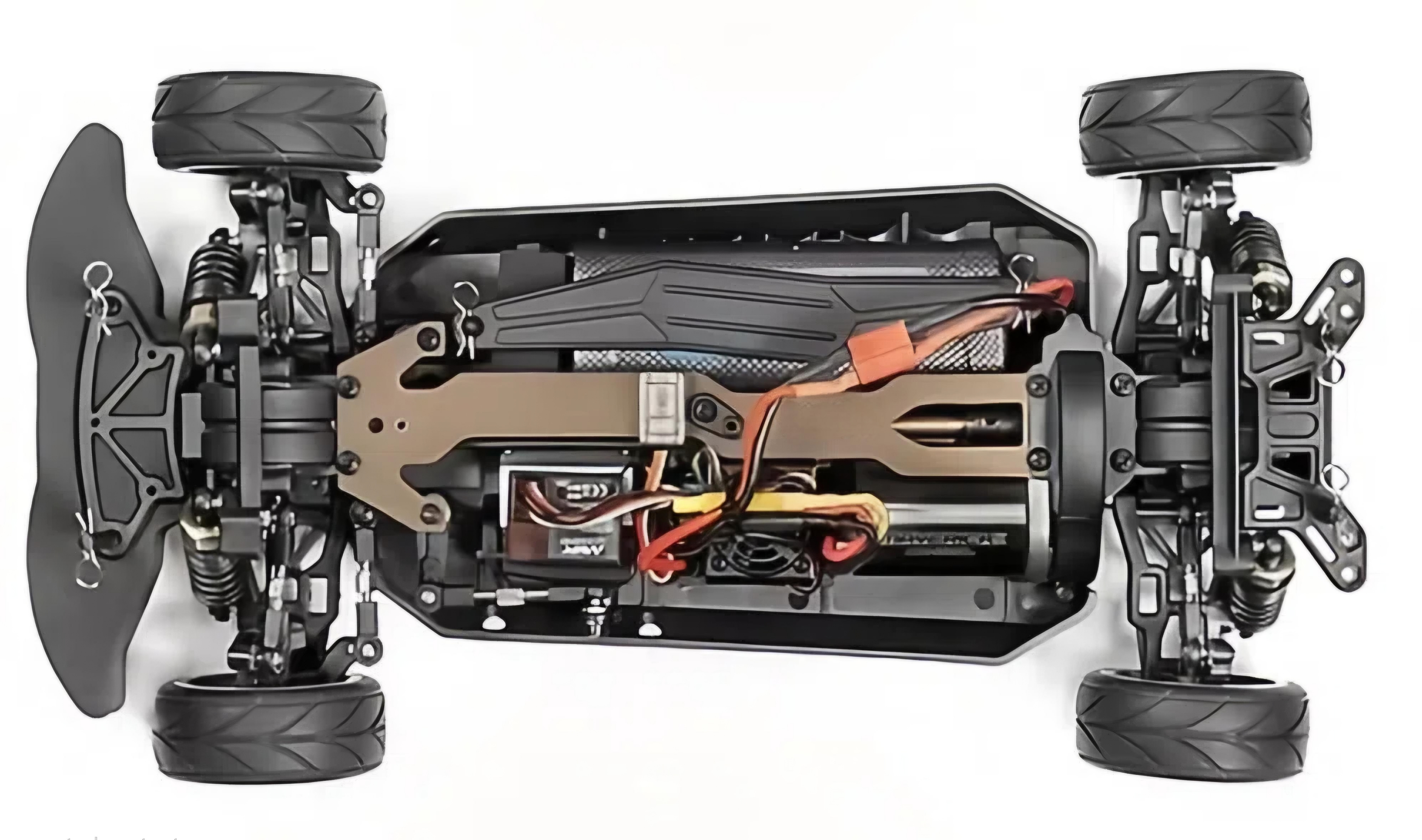}}}
    \qquad
    \subfloat[\centering]{{ \includegraphics[width=5.5cm]{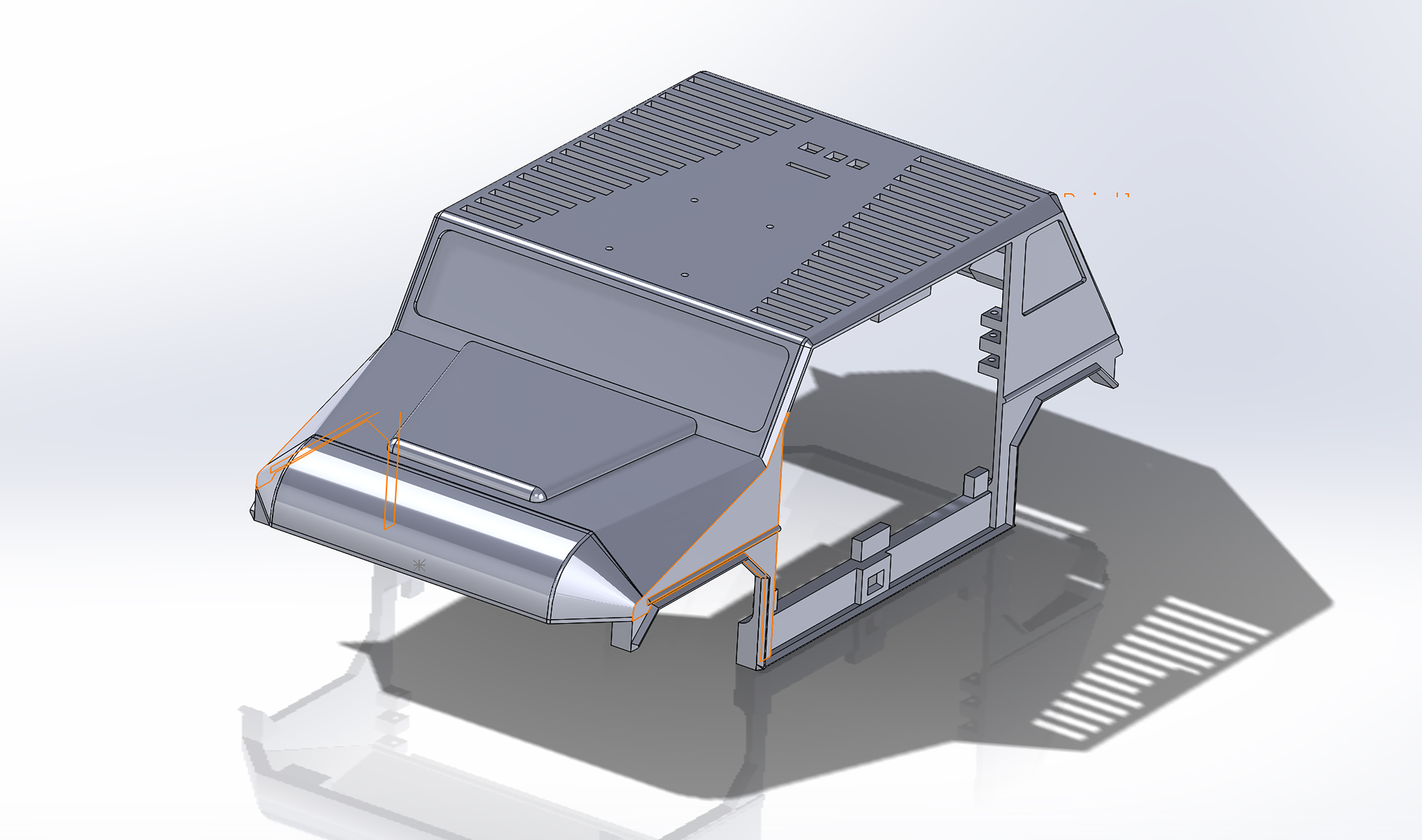}}} %
    \caption{Mechanical structure of robot: (a) Car chassis; and (b) Car body.} \label{fig2}
\end{figure}

The vehicle's chassis is adapted from a 1:10 scale RC car and modified to meet the specific requirements of the Auriga Robotics team as shown in Figure \ref{fig2}(a). Key modifications include an increased steering angle for better maneuverability and raising the car's overall height for improved clearance. The original DC motor was replaced with a stepper motor, offering enhanced precision and speed control, particularly at low velocities. Additionally, the chassis was reinforced to support heavier components such as larger batteries and control boards, ensuring durability and stability during operation.

\subsubsection{Car Body}
The car body and bumpers have been custom-designed to fulfill both functional and aesthetic needs as illustrated in Figure \ref{fig2}(b). The body structure not only provides a professional appearance but also allows for optimal airflow, which helps with cooling the internal components. Strategic mounting points for sensors and cameras have been incorporated into the design, facilitating seamless integration of distance sensors and vision systems required for autonomous navigation.

\newpage

\subsubsection{Power Transmission and Steering}
The power transmission system utilizes a single differential, with the rear axle functioning as the drive axle. Power is supplied by a stepper motor, which transmits torque through a gearbox and differential, ensuring smooth and efficient delivery to the wheels. This configuration provides reliable power distribution and consistent performance, allowing the vehicle to handle both high-speed and low-speed maneuvers effectively. The steering system, based on Ackermann steering geometry and driven by a servo motor, enables precise control of the vehicle’s direction, offering stable and responsive steering, especially during sharp turns and sudden directional changes. Together, these systems work in harmony to enhance the vehicle's overall performance, ensuring smooth power transmission and reliable control in dynamic driving environments.

\subsection{Electrical Design} 
{The electrical system of Auriga’s autonomous vehicle (AV) is built around a central control board and three auxiliary side boards, strategically positioned at the front, rear, and roof of the vehicle. These side boards facilitate the organized routing of sensor wires, ensuring neat and efficient connections throughout the system, while also reducing interference and signal loss. Power is supplied by a 5300 mAh 3-cell lithium polymer battery, which drives both the vehicle’s motor and its electronic components, providing consistent energy output for extended operation. Voltage regulation is critical in this system, with dedicated regulators maintaining voltages within safe operational limits to ensure the reliable performance of all equipment, preventing overheating or electrical damage during demanding tasks.}

\begin{figure}
    \centering
    \subfloat[\centering]{{ \includegraphics[width=5.5cm]{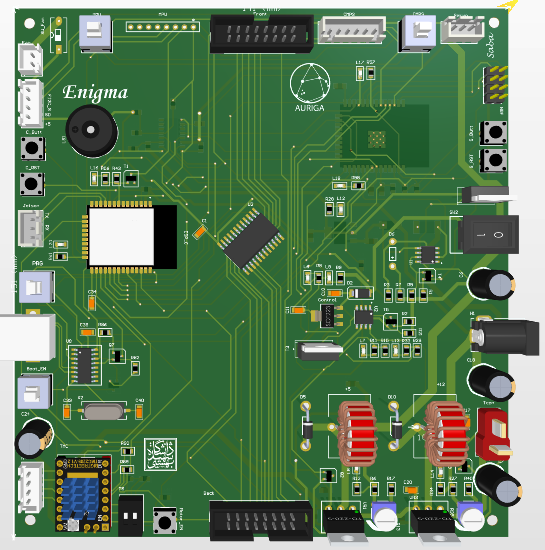}}} %
    \qquad
    \subfloat[\centering]{{ \includegraphics[width=5.5cm]{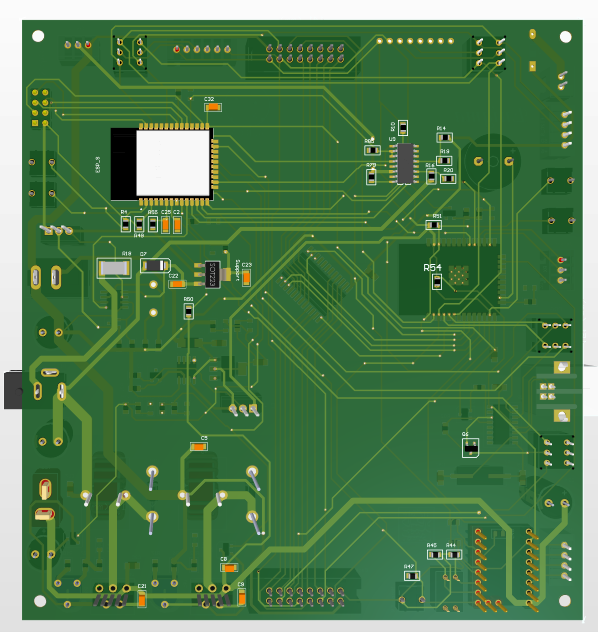}}} %
    \caption{Control board PCB layout: (a) Top Side; and (b) Bottom Side.} \label{fig3}
\end{figure}

\subsubsection{Control Board}
As shown in Figure \ref{fig3}, the main control board integrates two ESP32 microcontrollers that communicate via the UART protocol. One microcontroller handles control tasks like motor and sensor management, while the other oversees electrical monitoring, including real-time voltage and current measurements. To ensure proper voltage supply, the design incorporates switching and linear regulators. Two XL4016 switching regulators provide 10.5V and 5V to the Jetson Orin NX, servo motor, and various sensors. Additionally, two AMS1117-3.3V linear regulators supply power to the microcontrollers and some sensors. Data exchange between the microcontrollers and the Jetson Orin NX occurs through a Serial connection via UART. The entire PCB design and assembly were managed by the Electronics team, ensuring an optimized system for the AV.

\subsubsection{Sensors}
\begin{enumerate}
    \item{ \textbf{Camera}}
    The system is equipped with an IMX-219 camera module, featuring a wide 170-degree field of view. This camera was selected for its wide-angle lens, which provides an extensive view of the surrounding environment. The broad field of view is particularly advantageous for detecting objects and obstacles across a wide scene.
    
    \item{ \textbf{Range Sensors}}
    Time-of-flight laser-ranging sensors (VL53L1X) are mounted on the front, rear, and sides of the vehicle. These sensors provide high-speed and accurate distance measurements, which assist in obstacle detection and enhance the safety of the AV’s navigation. The sensor data complements the camera’s information to ensure precise environmental awareness.
    \item{ \textbf{IMU Sensors}}
    
    The vehicle employs two gyroscope sensors, the CMPS14 and MPU6050, to track its position during movement. These sensors measure angular changes, which contribute to localization, a key element in autonomous navigation. By combining data from both sensors, the system improves positional accuracy. Data is transmitted using the I2C communication protocol.
\end{enumerate}

\newpage

\section{Algorithm and Software Description}
\subsection{Embedded System Design}
In the development of autonomous vehicles, efficient control systems are vital for managing motor functions, sensor data acquisition, and communication between processors. To achieve this, an embedded system composed of two ESP32 microcontrollers, sensors, and corresponding drivers has been implemented. This system ensures seamless operation by dividing tasks across multiple components. The dual ESP32 microcontroller setup is designed to enhance task distribution and system efficiency. The main ESP32 takes charge of critical operations, such as running control algorithms (like PID), managing communication protocols, and collecting sensor data. On the other hand, the secondary ESP32 handles non-critical but essential tasks, including monitoring battery voltage via the ADC, displaying status information using NeoPixel LEDs, and enabling remote control functionality.

\begin{figure}
    \centering
    \subfloat[\centering]{{ \includegraphics[width=11cm]{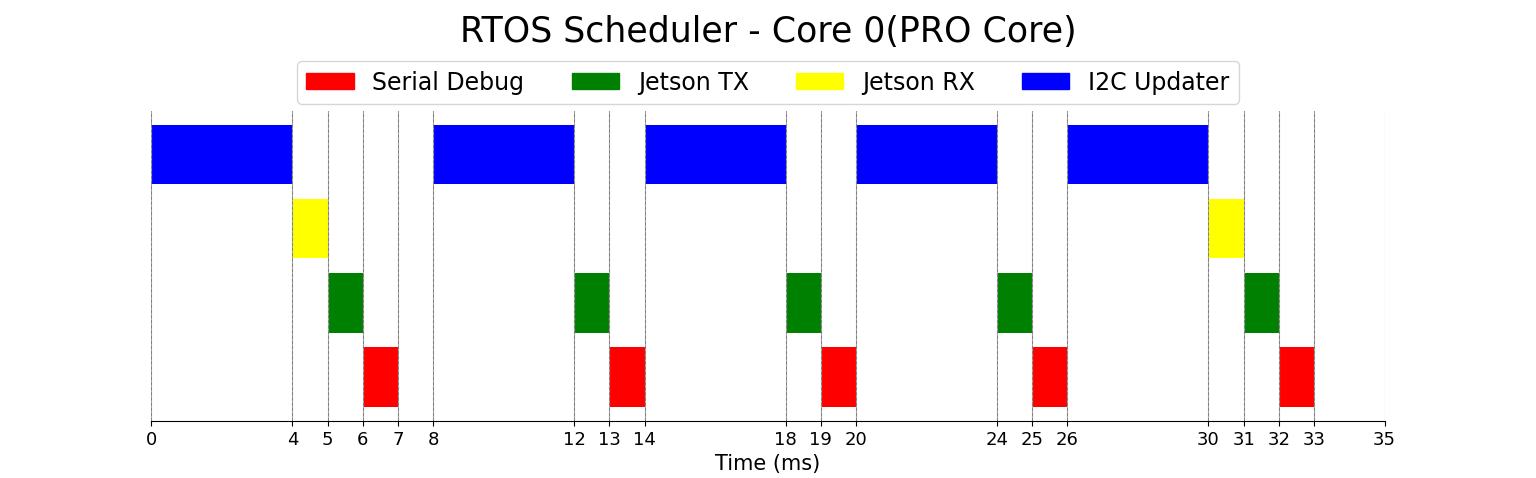}}} %
    \\
    \subfloat[\centering]{{ \includegraphics[width=11cm]{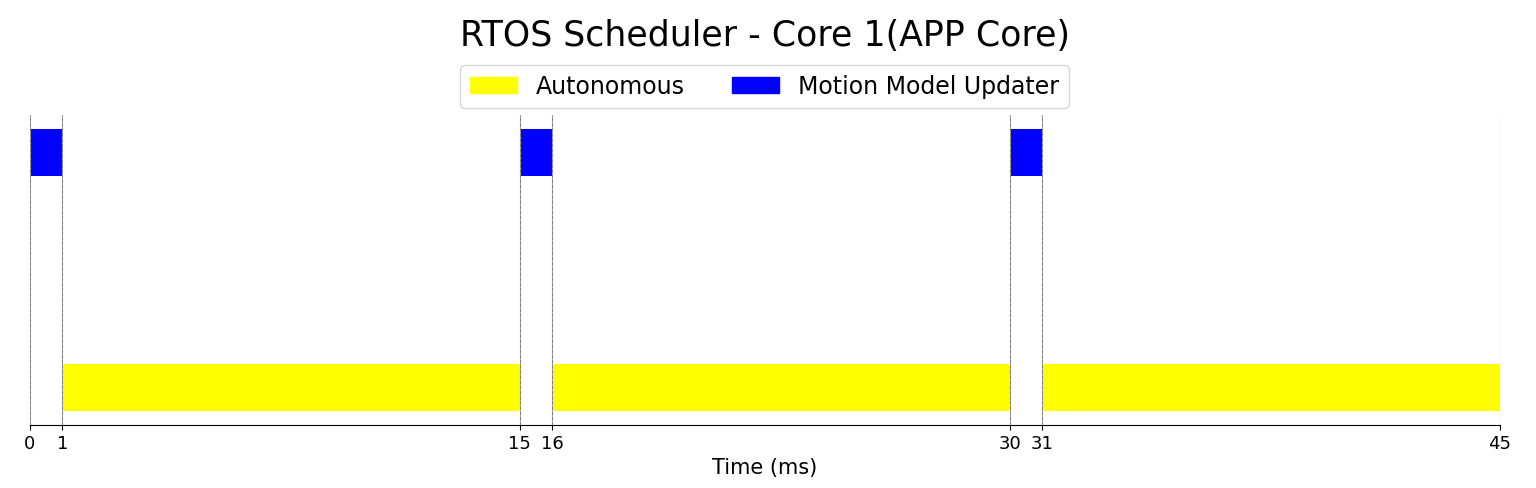}}} %
    \caption{RTOS task scheduling timeline: (a)Core 0; and (b)Core 1.} \label{fig4}
\end{figure}

\begin{figure}
\centering
\includegraphics[width=10cm]{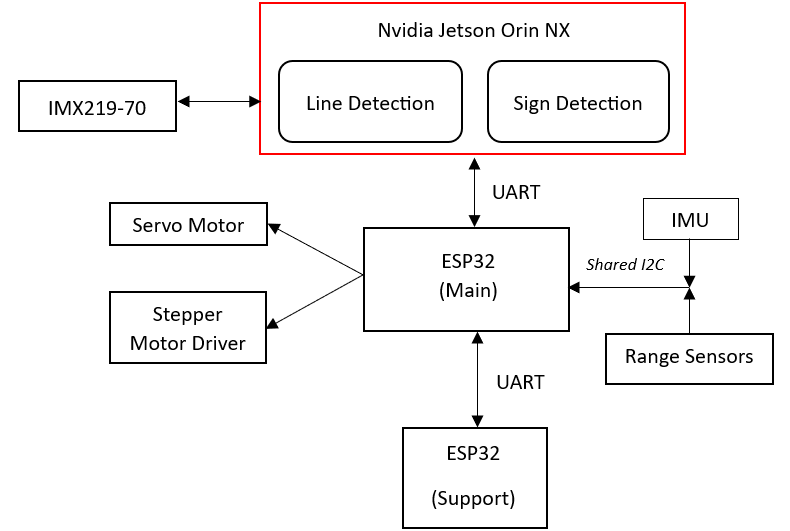}
\caption{Embedded system architecture.} \label{fig5}
\end{figure}

Task prioritization within the embedded system is managed using FreeRTOS, a real-time operating system that supports efficient multitasking, as shown in Figure \ref{fig4}. This ensures that higher-priority tasks are executed before lower-priority ones, allowing for optimized performance and stability in the vehicle’s operations.
As illustrated in Figure \ref{fig5}, Communication between the Jetson Orin NX and the main ESP32 is facilitated through the UART protocol, ensuring that data required for the control algorithms is transmitted reliably. The main and secondary ESP32s are also connected via the UART protocol, enabling smooth coordination between the two microcontrollers.

\vspace{-5pt}
\subsection{Computer Vision}
\subsubsection{Camera Calibration}
\begin{figure}[h]
    \centering
    \subfloat[\centering]{{ \includegraphics[width=4.5cm]{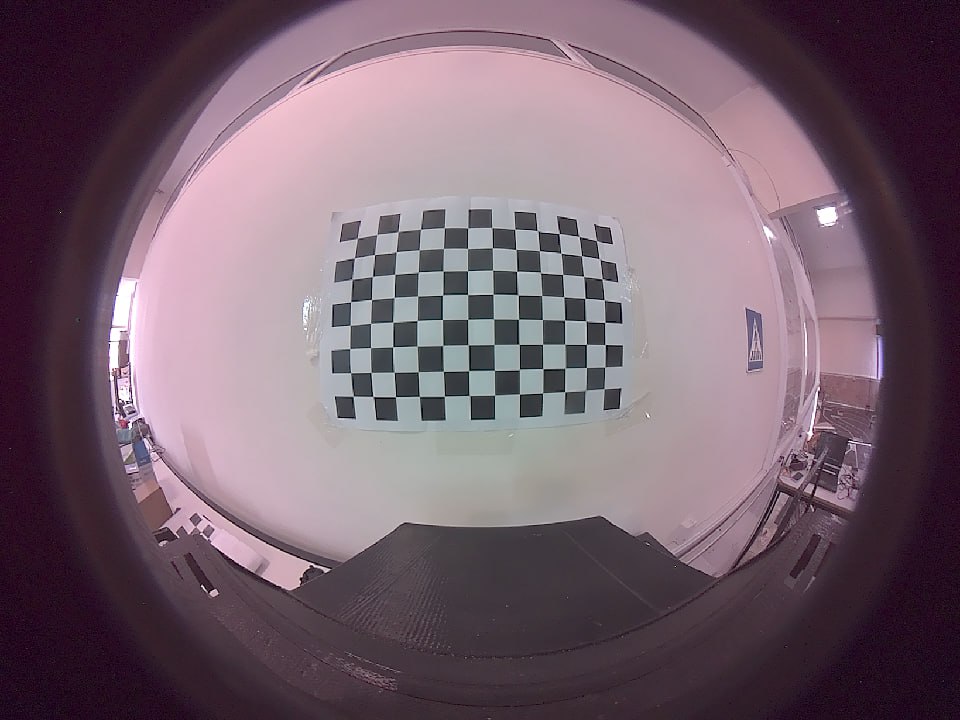}}} %
    \qquad
    \subfloat[\centering]{{ \includegraphics[width=4.5cm]{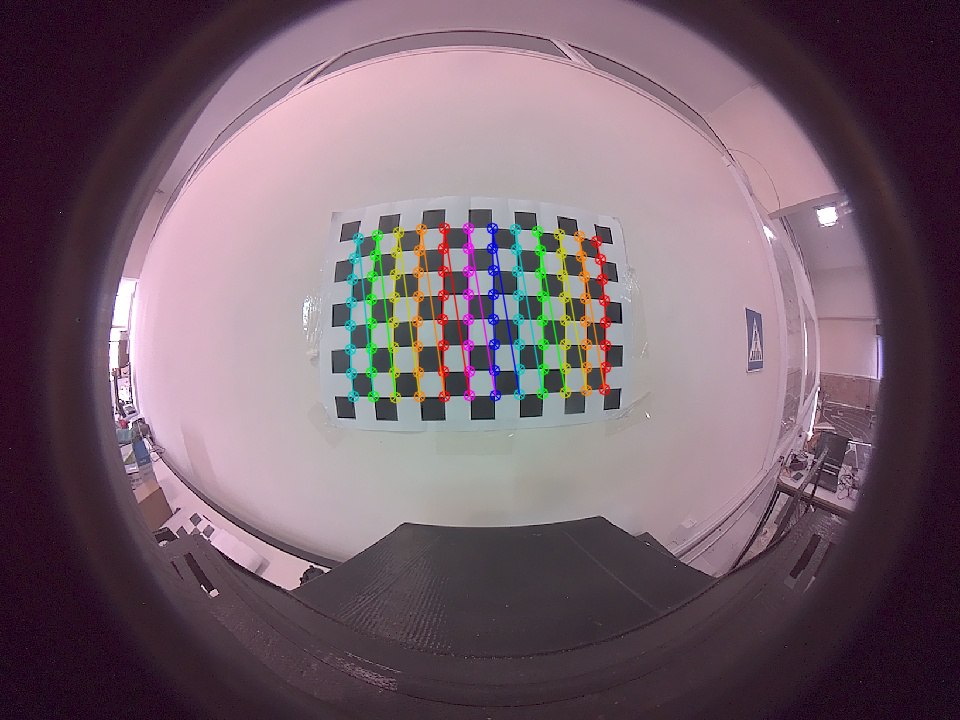}}} %
    \\
    \subfloat[\centering]{{ \includegraphics[width=4.5cm]{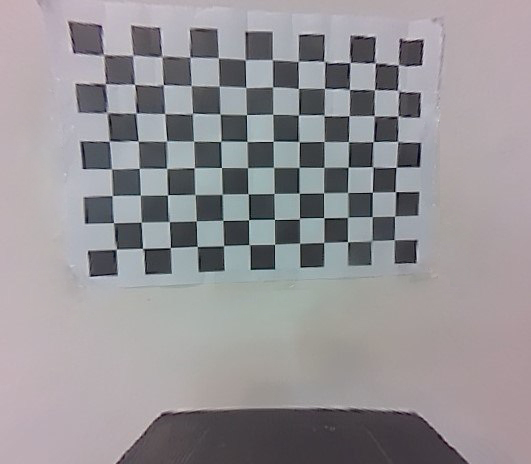}}} %
    \caption{The fisheye camera calibration process: (a) Capturing the initial image with the fisheye effect; (b) Detecting intersection points of the black and white squares on the chessboard; and (c) Outputting the calibrated image using the calculated coefficients to remove the distortion.} \label{fig6}
\end{figure}

Two main stages are involved in implementing a vision system for processing camera frames in a robot’s navigation: capturing and processing frames using GPU-accelerated techniques and publishing the processed data through shared memory for other systems. The process starts with camera initialization, where parameters like resolution, framerate, and calibration data are configured to correct lens distortions. A function is then applied to undistort the camera frame, leveraging GPU processing to ensure optimal performance, depicted in Figure \ref{fig6}.
Once the camera is ready, frames are captured using a method that establishes a video stream, and these captured frames are passed through a function to correct distortions, ensuring geometrically accurate images for further processing.
After processing the frames, the data and related flags are communicated to other systems via shared memory. This real-time communication includes essential flags for camera masking, crosswalk detection, and sign detection, which improves overall system efficiency by avoiding large-scale data duplication between processes. The system also detects crosswalks by subscribing to ROS messages, updating shared memory flags when a crosswalk is detected, and allowing other robot subsystems to adjust behavior accordingly based on this data.

\newpage

\vspace{-18pt}
Finally, the undistorted frames are written to shared memory, and flags are updated to signal that the frame is ready for other systems like path planning or obstacle avoidance. A message is then published via ROS to notify other nodes that camera data is available. The system operates continuously, processing frames in a loop until terminated. Upon shutdown, it releases the video capture device and cleans up shared memory to ensure an orderly conclusion to the process.

\subsubsection{Line Detection}
\begin{enumerate}
    \item {\textbf{Dataset}}\\
    For the line detection task, two datasets were utilized: one during the pretraining phase and another for the final training stage. Initially, the Tusimple dataset was employed for pretraining, providing a foundational understanding of road line detection \cite{DBLP:journals/corr/abs-2005-08630}. For the final phase, a custom, handmade dataset was created, specifically tailored to the conditions and requirements of the autonomous vehicle.
    The handmade dataset consists of approximately 39,150 images, which were augmented to enhance diversity and robustness. Ambient noise was introduced to simulate real-world conditions more accurately, improving the model’s capacity to handle challenging environments. All images from the dataset were included in the training set, and real-time evaluation during the vehicle’s operation ensured optimal system performance in real-world driving conditions.\\

    \item {\textbf{Network}}\\
    A key computer vision task for the autonomous vehicle is line detection, achieved through the use of PiNet, an instance segmentation model that was customized to meet the speed requirements of the vehicle \cite{DBLP:journals/corr/abs-2002-06604}. PiNet operates by isolating road lines from raw video frames in real time, allowing the vehicle to effectively detect and follow lanes.
    To enhance processing speed, the architecture of PiNet was modified by reducing the number of hourglass blocks from four to two. Although this adjustment led to a slight decrease in accuracy, the increased processing speed enabled faster vehicle operation while maintaining reliable lane detection. This trade-off between speed and accuracy was essential for real-time performance, which is critical in autonomous vehicle systems.

    \vspace{-8pt}
    \begin{table}
    \centering
    \caption{Test results on mx350}\label{tab1}
    \begin{tabular}{|c|c|c|}
    \hline
     & fps &  Parameters \\
    \hline
    Our model & 25 & 3.36M\\
    PiNet new & 18 & 4.06M\\
    PiNet & 26 & 4.39M\\
    \hline
    \end{tabular}
    \end{table}

    The original PiNet model, while optimized for speed, was primarily designed for detecting straight lines and demonstrated limitations in detecting curves. Customizing PiNet improved its ability to detect both straight and curved lines, aligning the model’s functionality with the unique challenges presented by the vehicle's operational environment.\\

    \item {\textbf{Post Process}}\\
    To perform the task of detecting lines and navigating concerning them, two primary components are required: first, the training of a neural network to detect points along a semantic line, and second, the post-processing of the network's output. This section provides a high-level explanation of the post-processing phase.
    The neural network outputs the coordinates of all detected points, which are obtained from the robot’s camera, aligned parallel to the road and its lines, as illustrated in Figure \ref{fig7}. However, to measure the geometry of these detected lines accurately, a top-down perspective is necessary, commonly referred to as Bird's Eye View (BEV), depicted in Figure \ref{fig8}.

    \begin{figure}
    \centering
    \includegraphics[width=7cm]{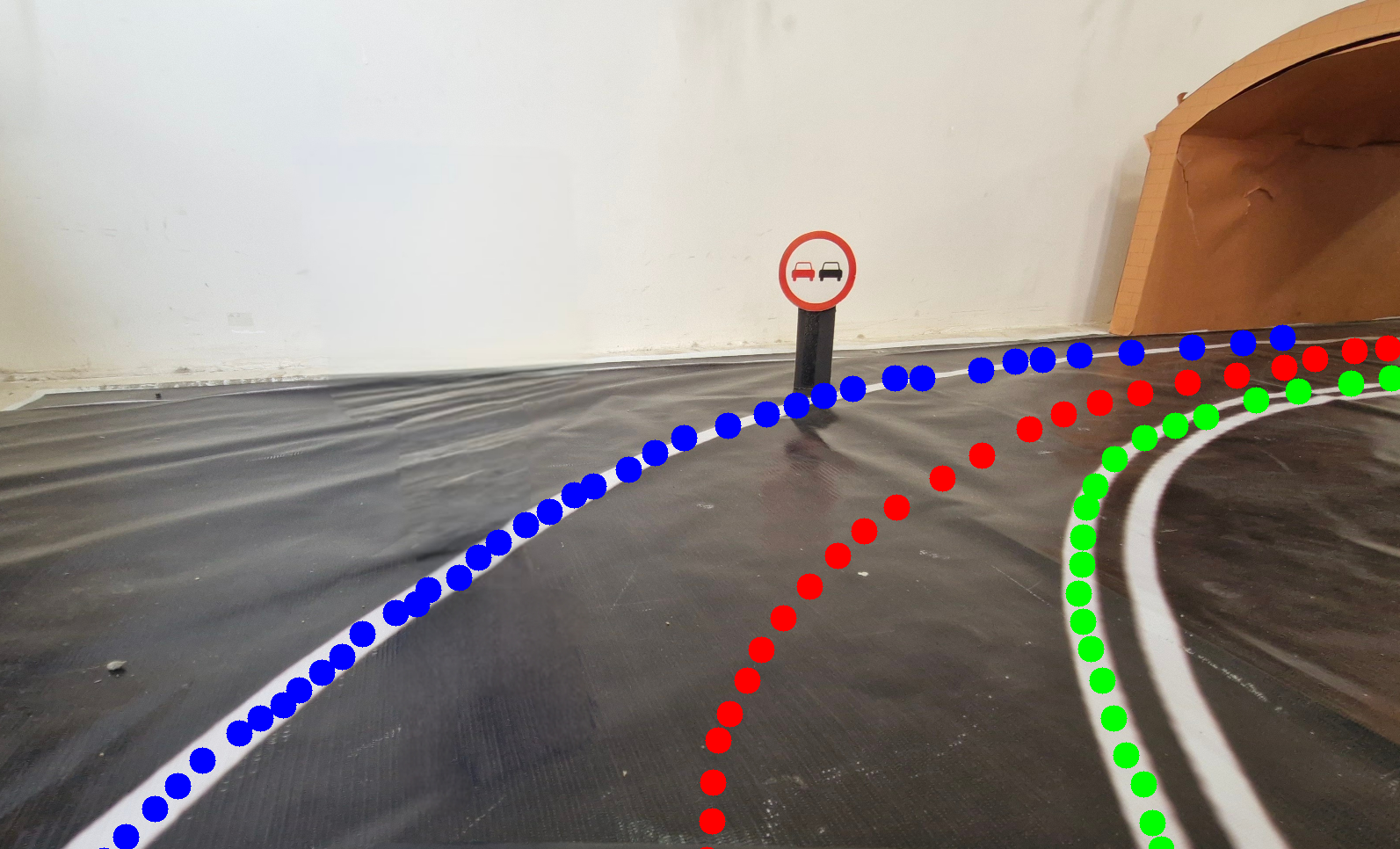}
    \caption{Output of the network for detecting semantic lane lines.} \label{fig7}
    \end{figure}
    
    For simplification, there is no need to transform the entire frame. Instead, by calculating the pixel-to-meter ratio for the camera, it is possible to transform only a portion of the frame. This portion’s length corresponds to 1.5 times the maximum length of the road ahead, while the width of the transformed frame may vary depending on factors such as camera placement.
    The robot requires two key pieces of information to navigate effectively: the distance from the lines and the curvature or angle of the lines. The initial step involves identifying the left and right lane markings. Once detected, a polynomial—quadratic or linear, depending on the points—is fitted to these markings. To compute the car’s distance from the lines and its angle relative to the road, a "middle line" is introduced. The middle line can be computed in one of three ways. If both lines are detected, the middle line is determined by averaging them. If only one line is detected, the middle line is created by shifting the detected line inward. When no lines are detected, the middle line is estimated using data from previous frames.

    \begin{figure}
    \centering
    \includegraphics[width=6cm]{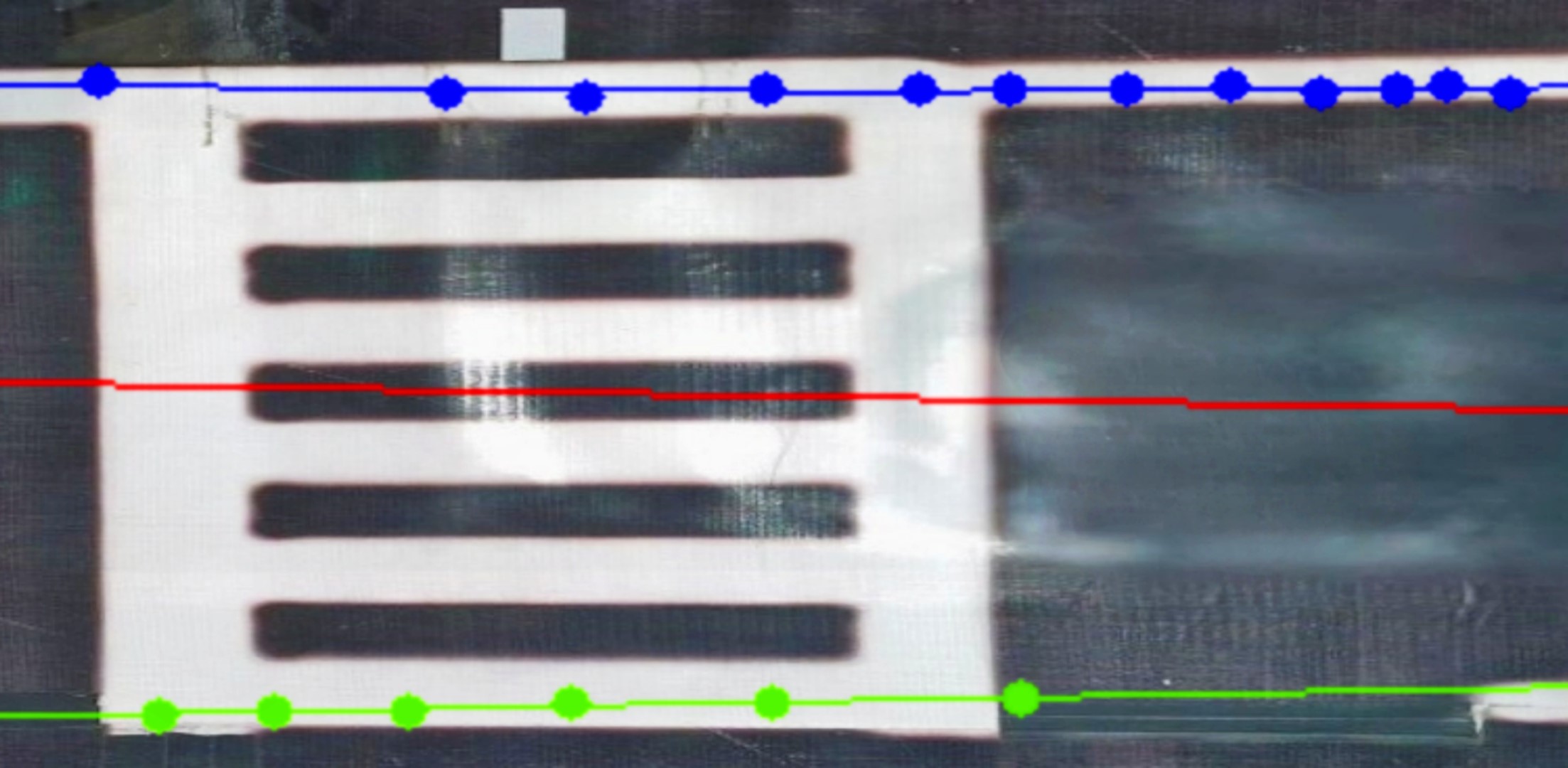}
    \caption{Bird eye view (BEV) of the captured frame.} \label{fig8}
    \end{figure}
    
    \vspace{-5pt}
    The middle line is then used to calculate the necessary navigation data. The distance is computed as the horizontal distance between the lowest point of the middle line and the center of the frame, while the angle is obtained by calculating the derivative of the middle line at this lowest point. To ensure reliability, outlier points that deviate significantly from the mean are filtered out, thereby preserving meaningful and consistent navigation data.\\

\end{enumerate}

\newpage

\subsubsection{Sign Detection}
\begin{enumerate}
    \item {\textbf{Dataset}}\\
    The dataset for sign detection is composed of two key components: synthetic data and real-world data. Synthetic data was generated to simulate real-world conditions, incorporating variations in lighting, angles, and obstructions. This approach enabled the creation of a controlled environment featuring specific road signs and backgrounds, allowing the model to train on a wide range of scenarios and edge cases that may be difficult to capture in natural settings. In addition to synthetic data, a real-world dataset was collected by capturing images in various road environments, ensuring diversity in sign types, road conditions, and lighting variations. This dataset included a diverse range of road signs and environmental conditions, further enhancing the robustness of the model and its ability to generalize to different real-world situations, making it highly adaptable to unpredictable driving environments.\\
    
    \item {\textbf{Network}}\\
    The object detection for signs was performed using the YOLO (You Only Look Once) architecture \cite{yolov8_ultralytics}. Initially, YOLOv8n (nano) was implemented due to its high frames per second (FPS) performance, achieving 45 FPS. However, it suffered from frequent misclassifications, leading the team to switch to YOLOv8s (small). While YOLOv8n offered speed, it lacked the accuracy required for reliable sign detection, with an accuracy rate of 0.82.

    \vspace{-8pt}
    \begin{table}
    \centering
    \caption{Test results on A100}\label{tab2}
    \begin{tabular}{|c|c|c|c|}
    \hline
    Model & mAP & Speed & Parameters \\
    \hline
    YOLOv8n & 18.4 & 1.21ms & 3.5M\\
    YOLOv8s & 27.7 & 1.40ms & 11.4M\\
    \hline
    \end{tabular}
    \end{table}
    
    In contrast, YOLOv8s delivered significantly improved detection accuracy at 0.95, albeit at a slower rate of 20 FPS. This trade-off between speed and accuracy made YOLOv8s a better fit for the system’s requirements. The performance of the system was thoroughly tested and evaluated based on detection accuracy, FPS, and distance measurement precision, ensuring that the vehicle could reliably detect road signs in real time.\\

    \item {\textbf{Post Process}}\\
    The post-processing stage focused on accurately calculating the distance between the camera and the detected road signs. This step presented a challenge due to the fisheye distortion caused by the wide-angle lens of the IMX-219 camera. To address this distortion, a mathematical model was applied, which considered the camera's focal length and the pixel height of the object within the image.

\newpage

    To understand the basics of lens optics, as illustrated in Figure \ref{fig9}, the relationship between sensor dimensions, field dimensions, focal length, and distance to the field can be summarized by the following equation:

    \begin{equation}
    \frac{\text{Sensor dimension (mm)}}{\text{Focal length } f \text{ (mm)}} = \frac{\text{Field dimension}}{\text{Distance to field } d}
    \label{eq1}
    \end{equation}

    \vspace{5pt}
    
    When substituting object size for field size, the object height on the sensor (mm) can be calculated as:

    \vspace{-5pt}
    \begin{equation}
    \text{Object height on sensor} = \frac{\text{Sensor height} \times \text{Object height (pixels)}}{\text{Sensor height (pixels)}}
    \label{eq2}
    \end{equation}
    
    This equation helps estimate the object's height on the sensor by using both the sensor and object dimensions in pixels. Using the real object size instead of pixels, the object height on the sensor can be expressed using the following formula:

    \begin{equation}
    \frac{\text{Object height on sensor (mm)}}{\text{Focal length (mm)}} = \frac{\text{Real object size}}{\text{Distance to object}}
    \label{eq3}
    \end{equation}

    \vspace{5pt}
    By rearranging the ratio, we can compute the unknown values such as distance to the object. For example:

    \begin{equation}
    \text{Distance to object} = \frac{\text{Real object height} \times \text{Focal length (mm)}}{\text{Object height on sensor (mm)}}
    \label{eq4}
    \end{equation}
    
    \vspace{5pt}
    Finally, the real object height can be derived using this equation:

    \vspace{-5pt}
    \begin{equation}
    \text{Real object height} = \frac{\text{Distance to object} \times \text{Object height on sensor}}{\text{Focal length}}
    \label{eq5}
    \end{equation}

    \vspace{-12pt}
    
    \begin{figure}
    \centering
    \includegraphics[width=7.2cm]{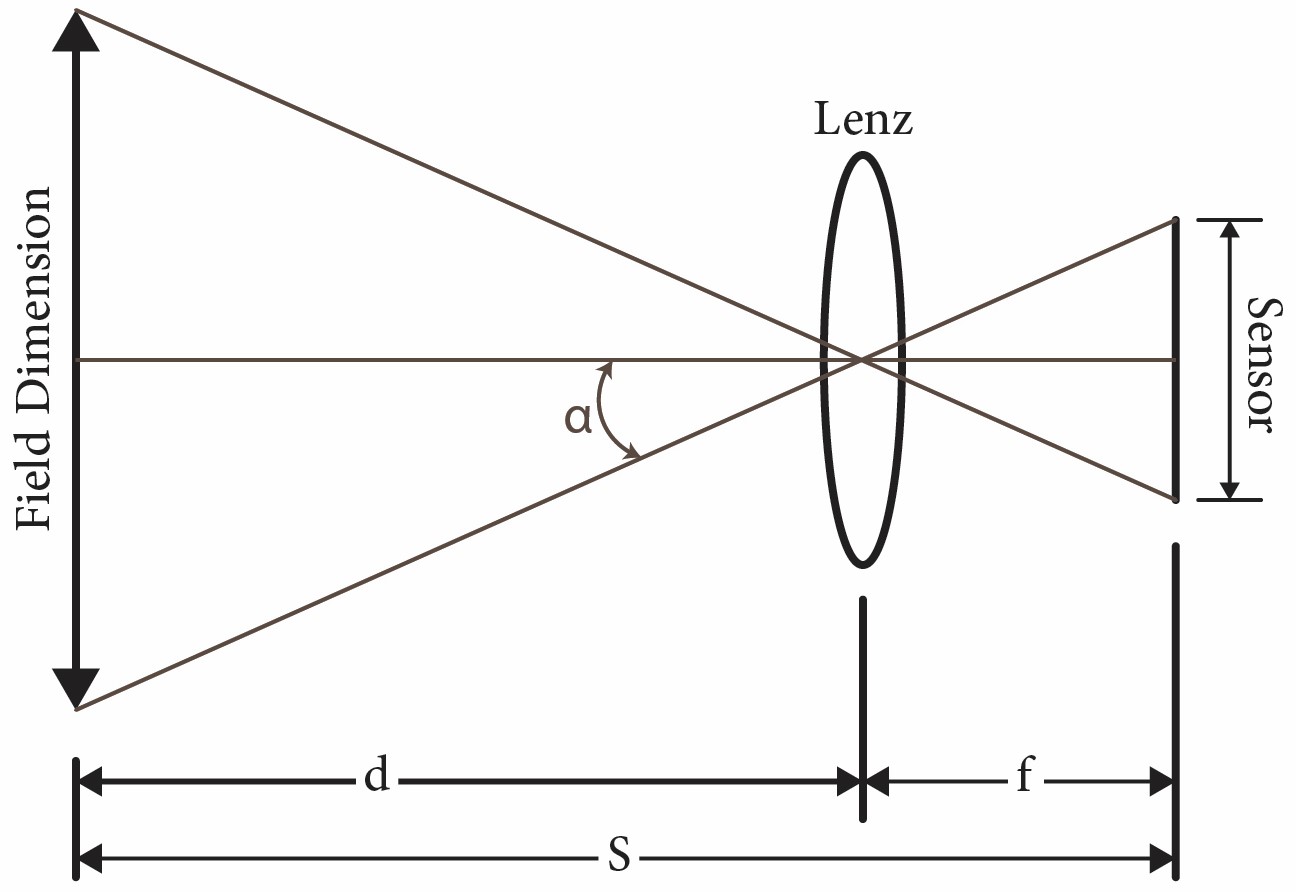}
    \caption{Distance measuring with focal length.} \label{fig9}
    \end{figure}

    \newpage
    
    In addition to this, machine learning techniques were utilized to refine the distance measurements further. A regression model was employed to find a correction coefficient and intercept that adjusted the computed distance. These adjustments shifted the distance to align with the road’s center (i.e., straight ahead), ensuring more precise distance measurements despite the camera’s distortion.\\

    \vspace{-10pt}
    
    \begin{figure}
    \centering
    \includegraphics[width=6cm]{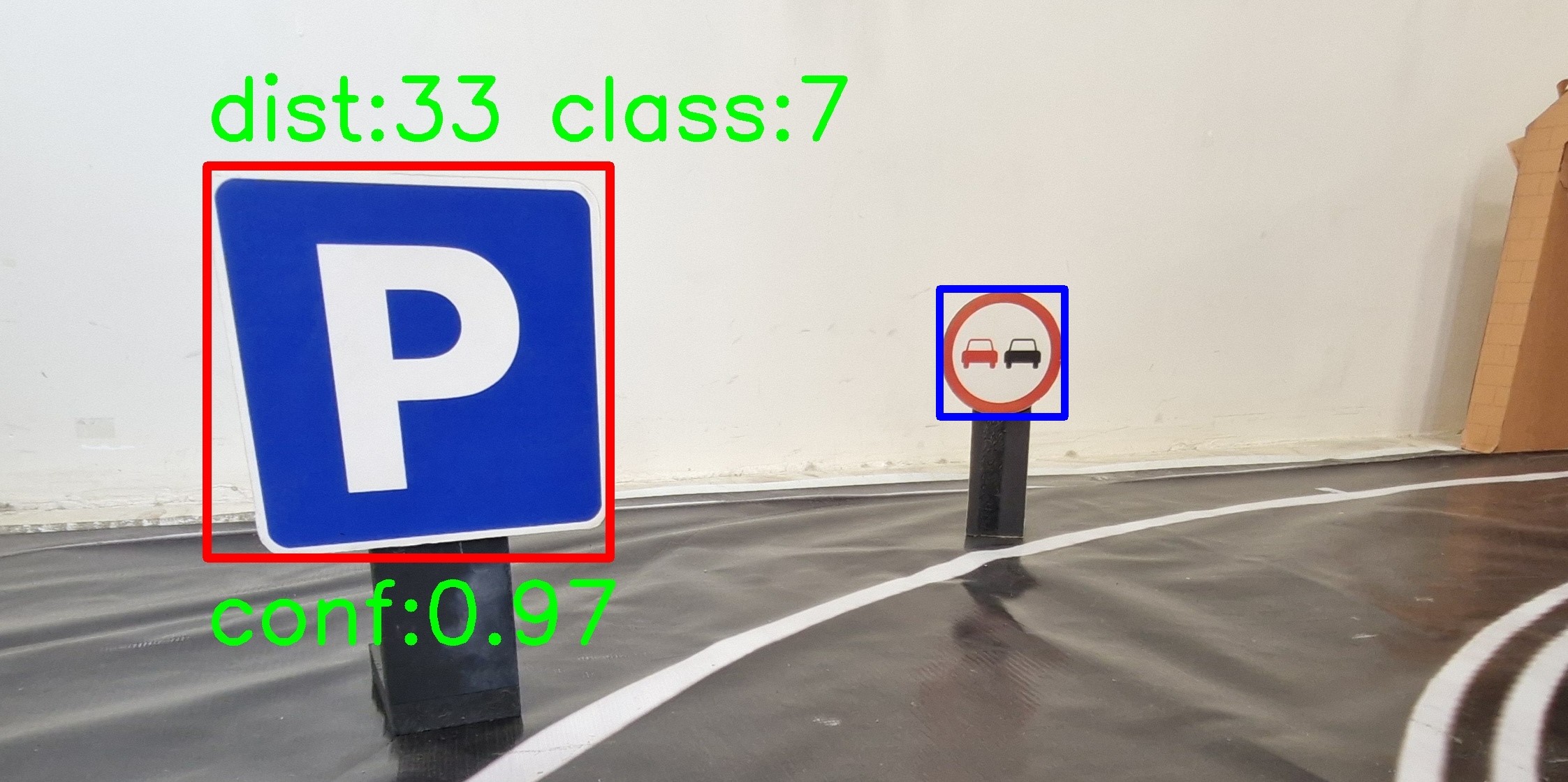}
    \caption{Sign detection output with bounding boxes, class, and nearest distance.} \label{fig10}
    \end{figure}
    
\end{enumerate}

\vspace{-35pt}

\subsection{Robot Operating System}
\begin{figure}
\centering
\includegraphics[width=11cm]{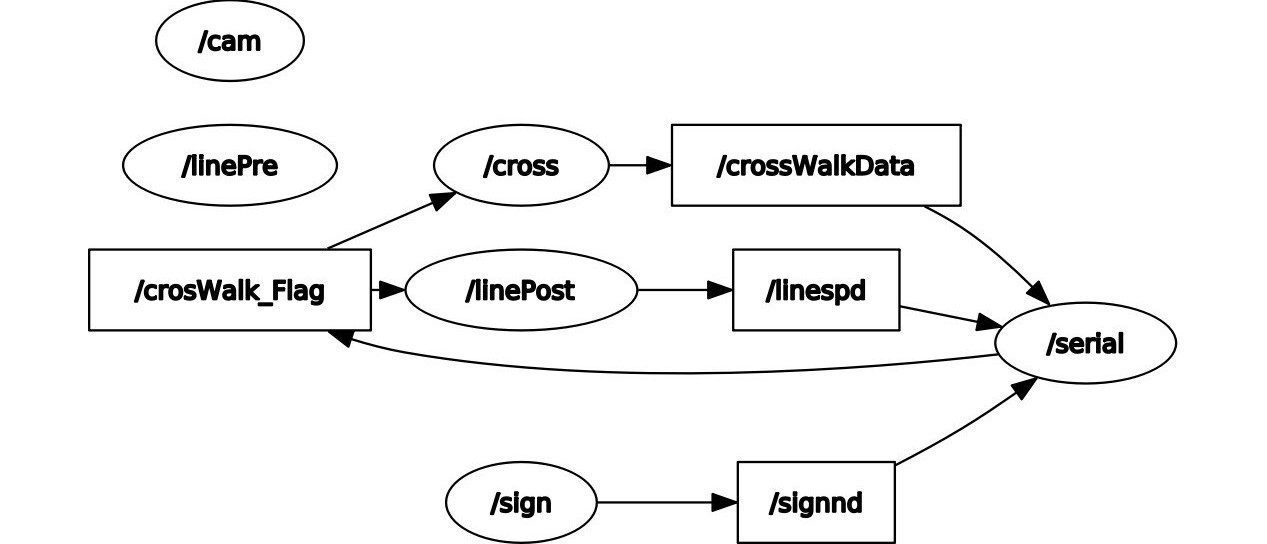}
\caption{RQT graph of task coordination and data flow.} \label{fig11}
\end{figure}

The Robot Operating System (ROS) plays a fundamental role in the design and functionality of the autonomous vehicle, providing a framework for efficient communication and coordination among the vehicle’s various computational modules \cite{inproceedings1}. ROS facilitates the seamless interaction between nodes that manage key tasks, such as computer vision, navigation, and sensor data processing, as illustrated in the RQT graph in Figure \ref{fig11}. At the core of the system, the Jetson NX processes camera outputs and calibrates visual data, ensuring it aligns with the vehicle’s navigation needs. Given the high frequency of visual data, traditional data-sharing methods are insufficient for real-time processing. To address this, Python’s shared memory is employed to enable faster communication between nodes, improving the vehicle’s responsiveness to changes in its environment.

\newpage

The ROS architecture supports simultaneous operations across different nodes, each dedicated to specific tasks like lane detection, traffic sign recognition, and crosswalk identification. Data from these processes is published to relevant ROS topics, which are used for navigation and decision-making. Task parallelization ensures efficient processing and minimizes delays in the system, allowing the vehicle to react swiftly to dynamic road conditions.

\subsection{Localization}
\begin{figure}
\centering
\includegraphics[width=8cm]{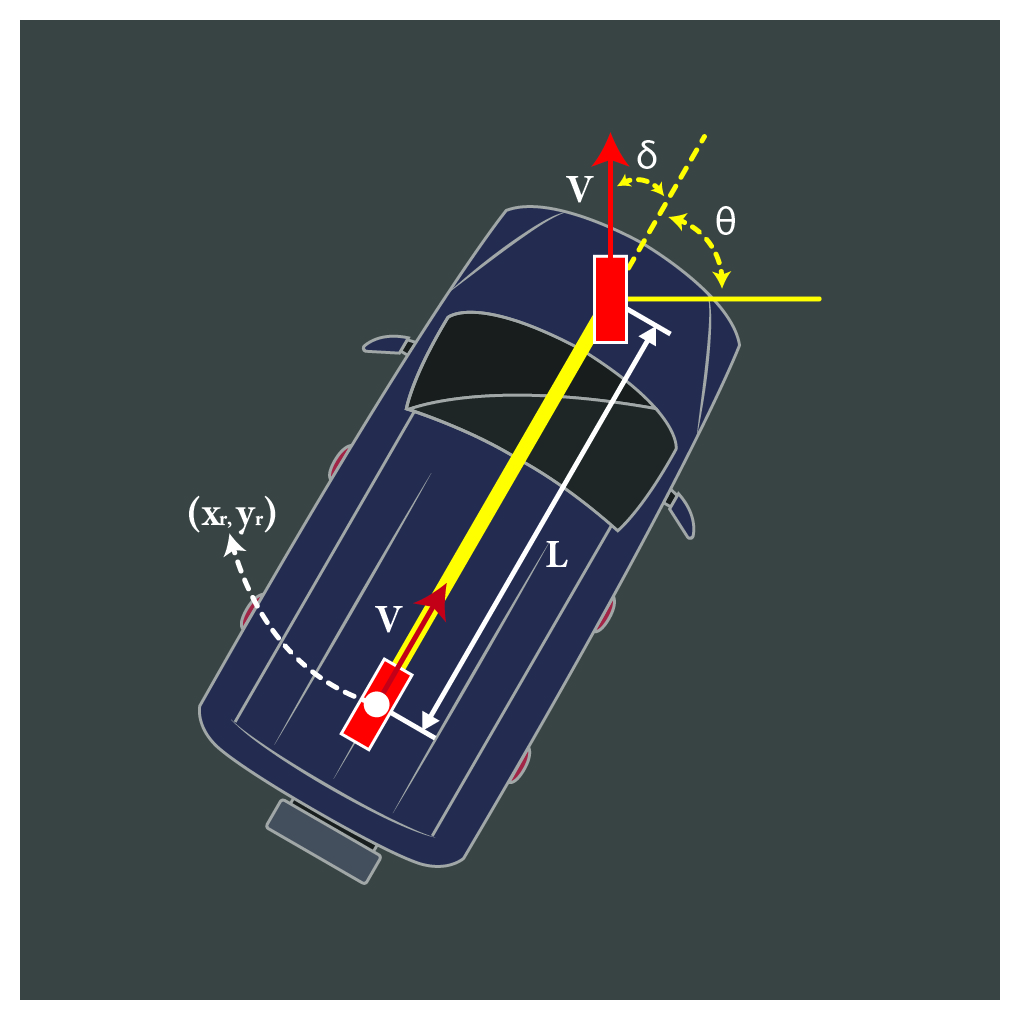}
\caption{Bicycle model of the robot’s kinematics.} \label{fig12}
\end{figure}

Accurate localization is crucial for an autonomous vehicle to determine its position and effectively perceive its surrounding environment. In this project, the vehicle’s physical model is simplified using the Bicycle Model, which is a common approach for vehicle kinematics. This model, as depicted in Figure \ref{fig12}, utilizes the vehicle’s speed and steering angle as inputs to compute three essential components for positioning: X (longitudinal position), Y (lateral position), and $\theta$ (orientation angle).

The longitudinal and lateral position of the vehicle can be expressed as:

\begin{equation}
\dot{x}_r = v \cos \theta
\label{eq6}
\end{equation}

\begin{equation}
\dot{y}_r = v \sin \theta
\label{eq7}
\end{equation}

\newpage

Finally, the vehicle's orientation is updated using a method \cite{khalili2024technicalreportmobilemanipulator} that combines IMU measurements and the following equation, where $\delta$ represents the steering angle and Length refers to the distance between the front and rear axles:

\vspace{-5pt}
\begin{equation}
\dot{\theta} = \frac{v \tan \delta}{\text{Length}}
\label{eq8}
\end{equation}

These components are calculated within a coordinate system established at the vehicle's initial position and are continuously updated in real time as the vehicle navigates through its environment. The motion equations, based on the vehicle’s rear axle, ensure accurate localization and enable smooth and responsive control of the vehicle as it moves \cite{7995816}.

\subsection{Mapping}
\begin{figure}
\centering
\includegraphics[width=5cm]{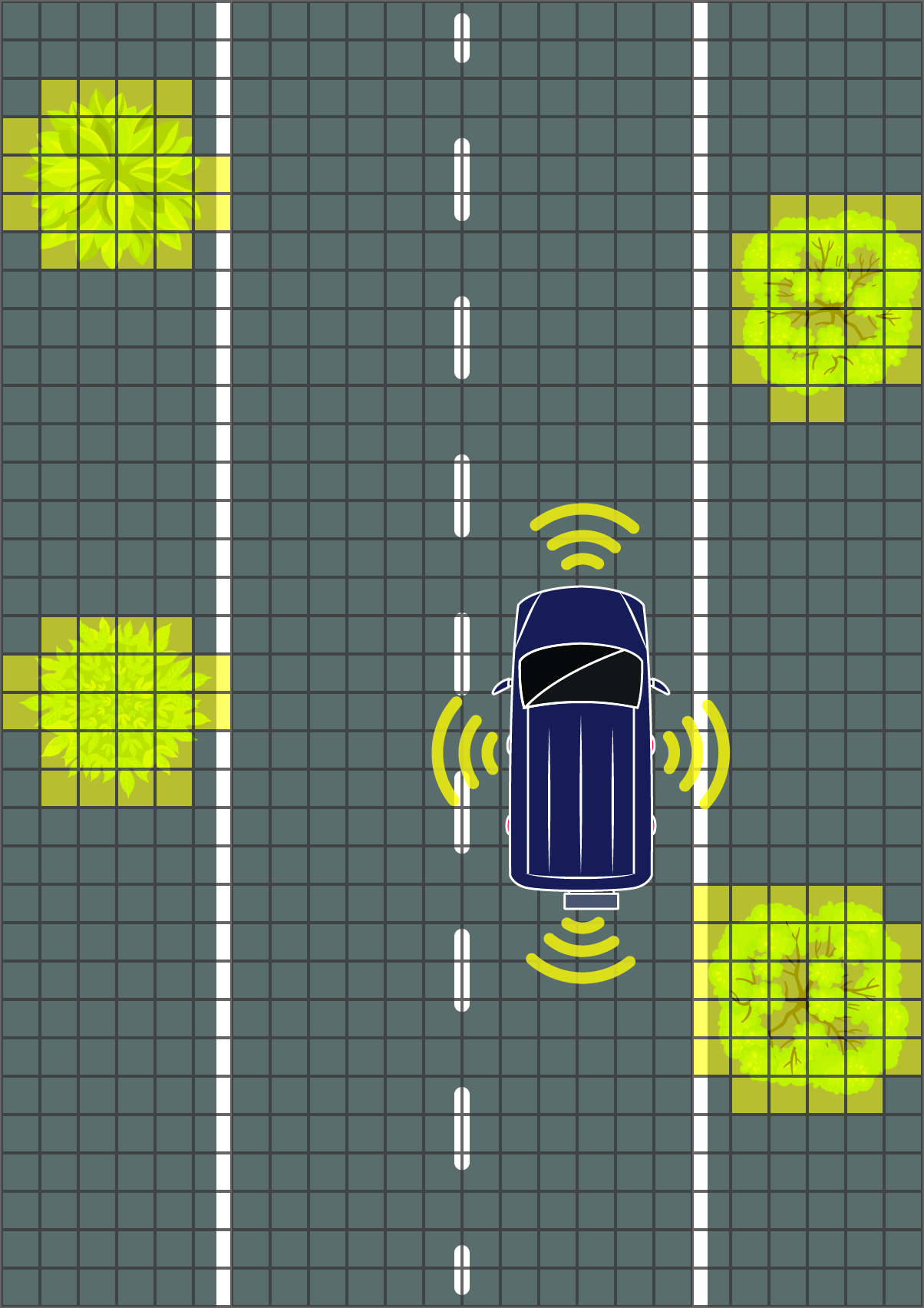}
\caption{Real-time mapping.} \label{fig13}
\end{figure}

The Auriga autonomous vehicle is equipped with the ability to generate real-time maps of its environment, including the road and any obstacles along the way. This is achieved using distance sensors mounted on the vehicle’s body, in combination with the localization coordinates described in the previous section. The surrounding environment is divided into uniform grid cells, which are continuously updated as the vehicle moves, as shown in Figure \ref{fig13}.
To minimize errors from the distance sensors, an algorithm is employed to validate sensor data. This algorithm estimates the state of each grid cell once a certain confidence margin is reached, ensuring more reliable and accurate mapping. By continuously refining this map, the vehicle can navigate safely and efficiently through dynamic environments.

\newpage

\subsection{Navigation}
Auriga autonomous vehicle employs several control algorithms to navigate and maintain lane positioning on the road. Each controller is chosen based on its specific strengths and operational flexibility.

\vspace{-5pt}
\begin{figure}
\centering
\includegraphics[width=8cm]{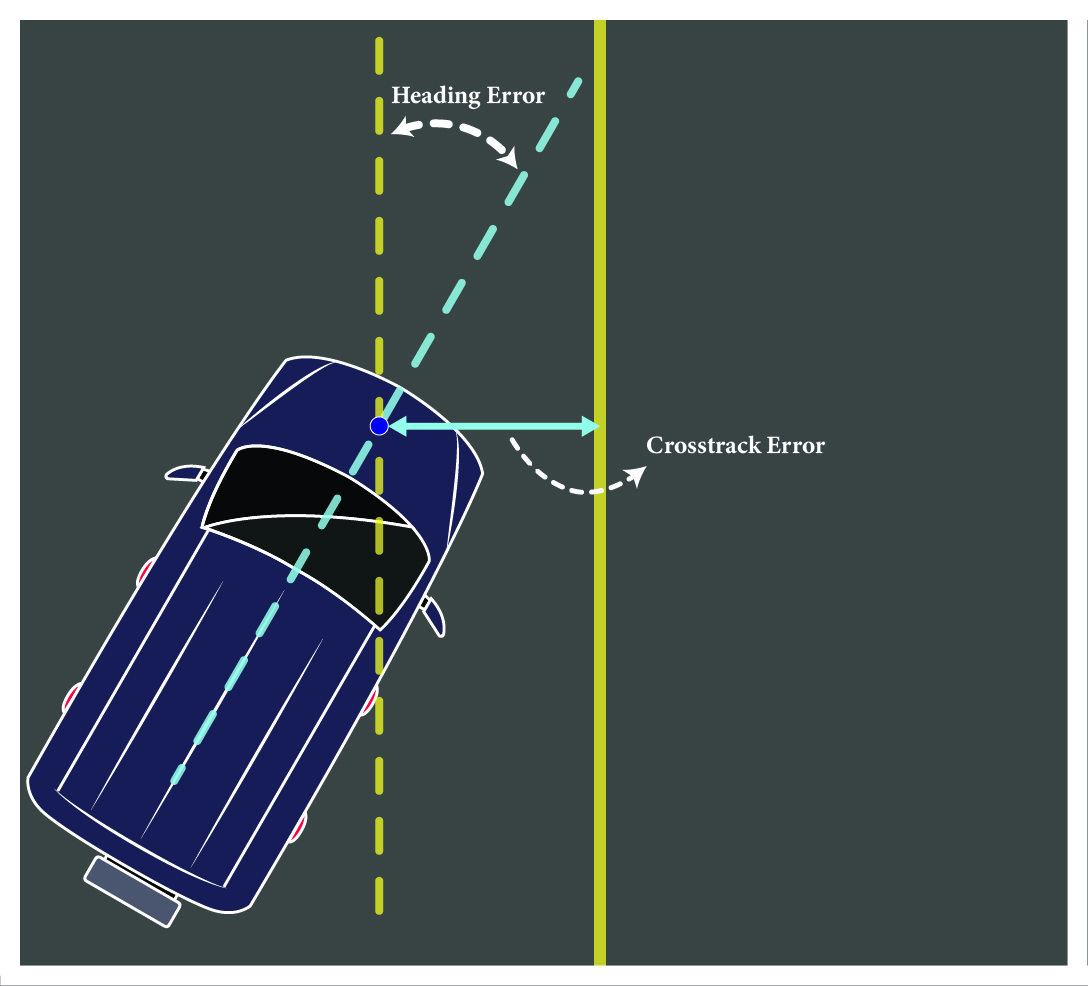}
\caption{Heading error (HE) and cross-track error (CE) for steering in navigation.} \label{fig14}
\end{figure}

\vspace{-25pt}
\subsubsection{Stanley Controller}
The primary algorithm for maintaining the vehicle's lane position adjusts the steering based on the vehicle's speed, the lateral distance from the road’s centerline, and the vehicle’s orientation relative to that line, as illustrated in Figure \ref{fig14}. The Stanley controller computes the steering angle as follows:

\begin{equation}
\text{Steer} = (k_{he} \times \text{Heading Error}) + \tan^{-1} \left( \frac{k_{ce} \times \text{Cross-track Error}}{v_f + k_{s}} \right) \label{eq9} 
\end{equation}

\vspace{10pt}
This method is highly adaptable to changes in speed, making it ideal for situations where the vehicle's velocity fluctuates during lane-keeping. One of the key advantages of this approach is its speed-sensitive control mechanism, ensuring smoother transitions when the vehicle accelerates or decelerates, crucial for maintaining stability while navigating within road lanes. This feature makes it the preferred method for continuous lane-keeping tasks \cite{article1}.

\subsubsection{PID Controller}
A backup system was implemented to support the primary steering control, with its parameters fine-tuned through extensive testing to ensure precise performance. The PID controller calculates the steering angle using the following equation:

\begin{equation}
\text{Steer} = (k_p \times \text{CE}) + (k_d \times \text{HE}) + \left( k_i \int \text{CE} \right)
\label{eq10}
\end{equation}

\vspace{10pt}
The proportional and integral components manage the vehicle's distance from the road’s centerline, while the derivative component adjusts steering based on the vehicle’s angle relative to that line. This approach ensures a balanced and responsive control of the vehicle’s steering \cite{article2}. In the vehicle’s final version, the PID controller is also used for speed regulation, where the P gain influences the curvature of the path and adjusts the vehicle’s speed accordingly, reducing it when necessary to safely navigate curves. Additionally, the P controller is integrated into the robot’s speed control system to manage obstacle avoidance. In cases where obstacles are detected, the obstacle controller takes higher priority, ensuring the vehicle safely maneuvers while maintaining optimal speed \cite{inproceedings2}.

\subsubsection{Pure Pursuit Controller}
Another algorithm tested on the vehicle yielded favorable results. The Pure Pursuit controller calculates the steering angle based on the distance of the vehicle's rear axle from the centerline of the road and its speed. The steering angle is computed as:

\vspace{5pt}
\begin{equation} \text{Steer} = \tan^{-1} \left( \frac{2 \times \text{Cross-track Error} \times \text{Length}}{k_{pp} \times \text{Speed}} \right) 
\label{eq11}
\end{equation}

\vspace{10pt}
This method is known for its robustness in path tracking, proving highly effective for vehicles navigating curved or complex paths. The advantage of the Pure Pursuit Controller is its ability to smoothly follow curved roads and track complex paths, even at higher speeds. Its reliance on geometric relationships between the vehicle and the path provides precise steering control, ensuring the vehicle stays centered within its lane \cite{6987787}.

\subsection{Decision Making}
The decision-making process of the Auriga autonomous vehicle is designed to handle various driving scenarios based on road conditions and detected signs. The vehicle continues to follow road lanes until a sign is detected, at which point it begins the appropriate operations based on the detected sign.

\subsubsection{Parks}

\begin{figure}
\centering
\includegraphics[width=6.5cm]{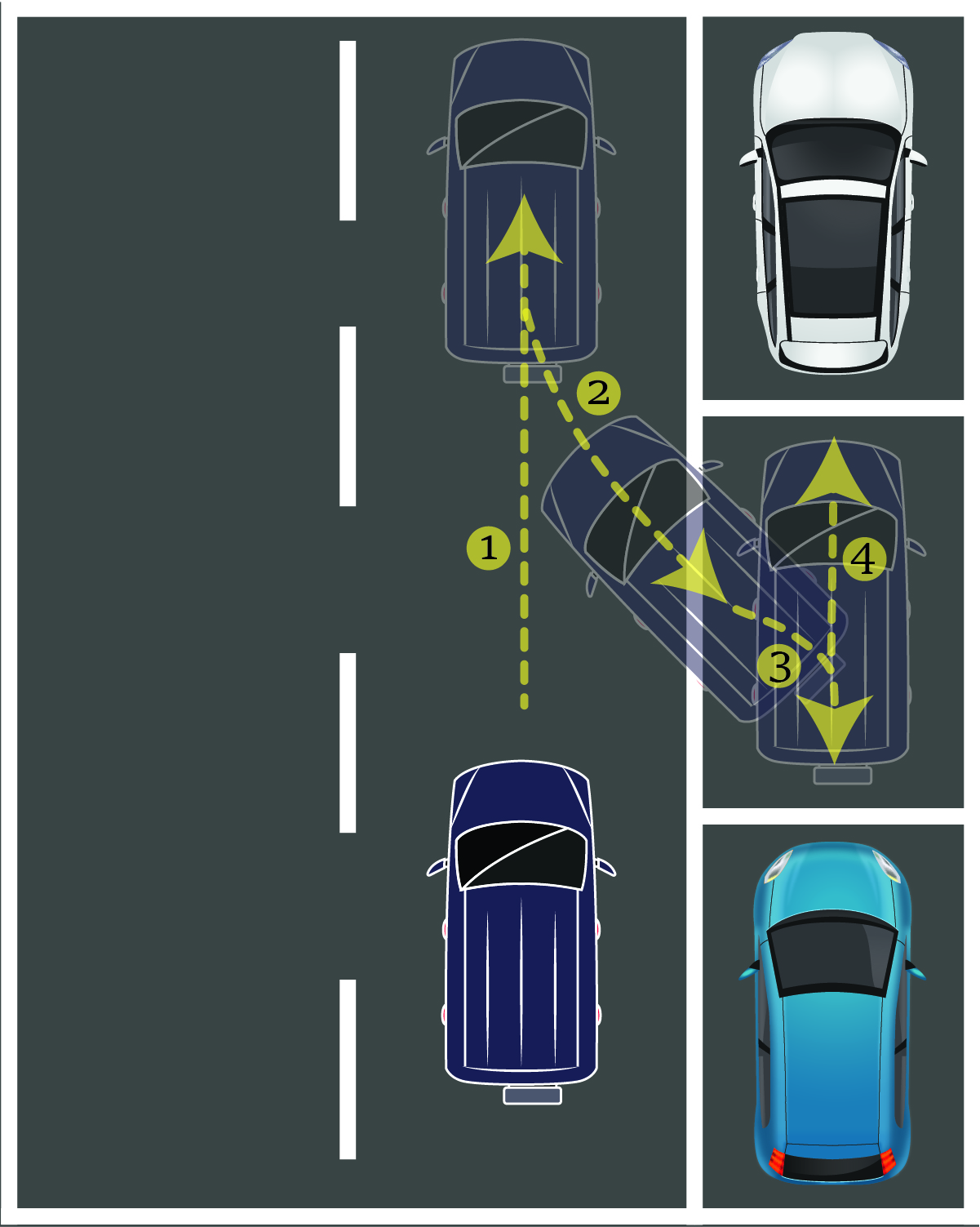}
\caption{Parallel parking maneuver.} \label{fig15}
\end{figure}

Upon detecting a parking sign, the vehicle initiates a sequence of actions governed by a Finite State Machine (FSM), as shown in Figure \ref{fig15}. The first step involves activating the hazard lights, followed by the use of distance sensors on both sides of the vehicle to identify a suitable parking space. Once a free space is detected, the vehicle proceeds to park. Throughout the parking process, the vehicle continuously monitors its surroundings using the distance sensors to avoid collisions with both static and dynamic obstacles. After completing the parking maneuver and ensuring all conditions for proper parking are met, the vehicle exits the parking mode and resumes normal navigation.

\subsubsection{Intersections}
\begin{figure}[h]
    \centering
    \subfloat[\centering]{{ \includegraphics[width=4cm]{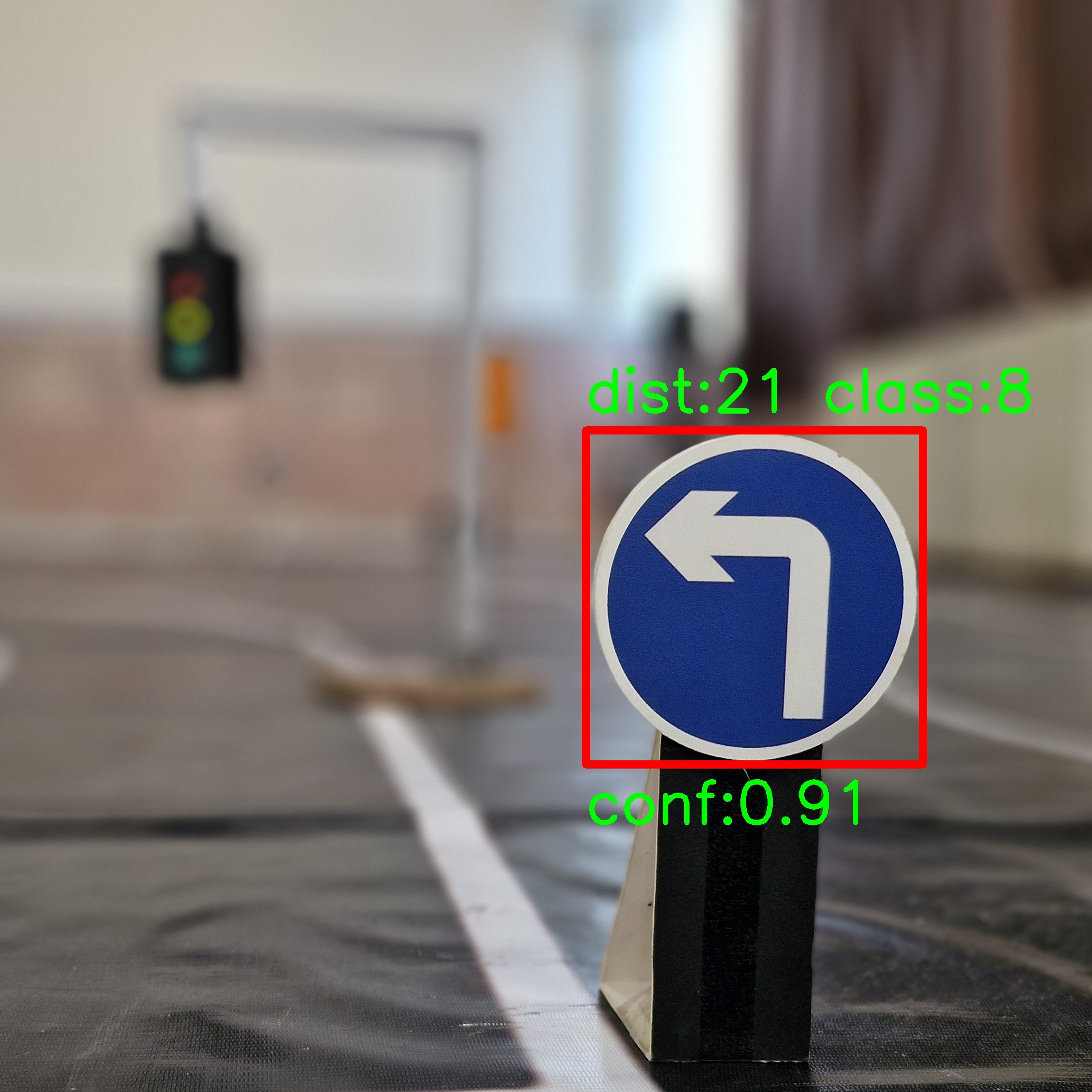}}} %
    \qquad
    \subfloat[\centering]{{ \includegraphics[width=4cm]{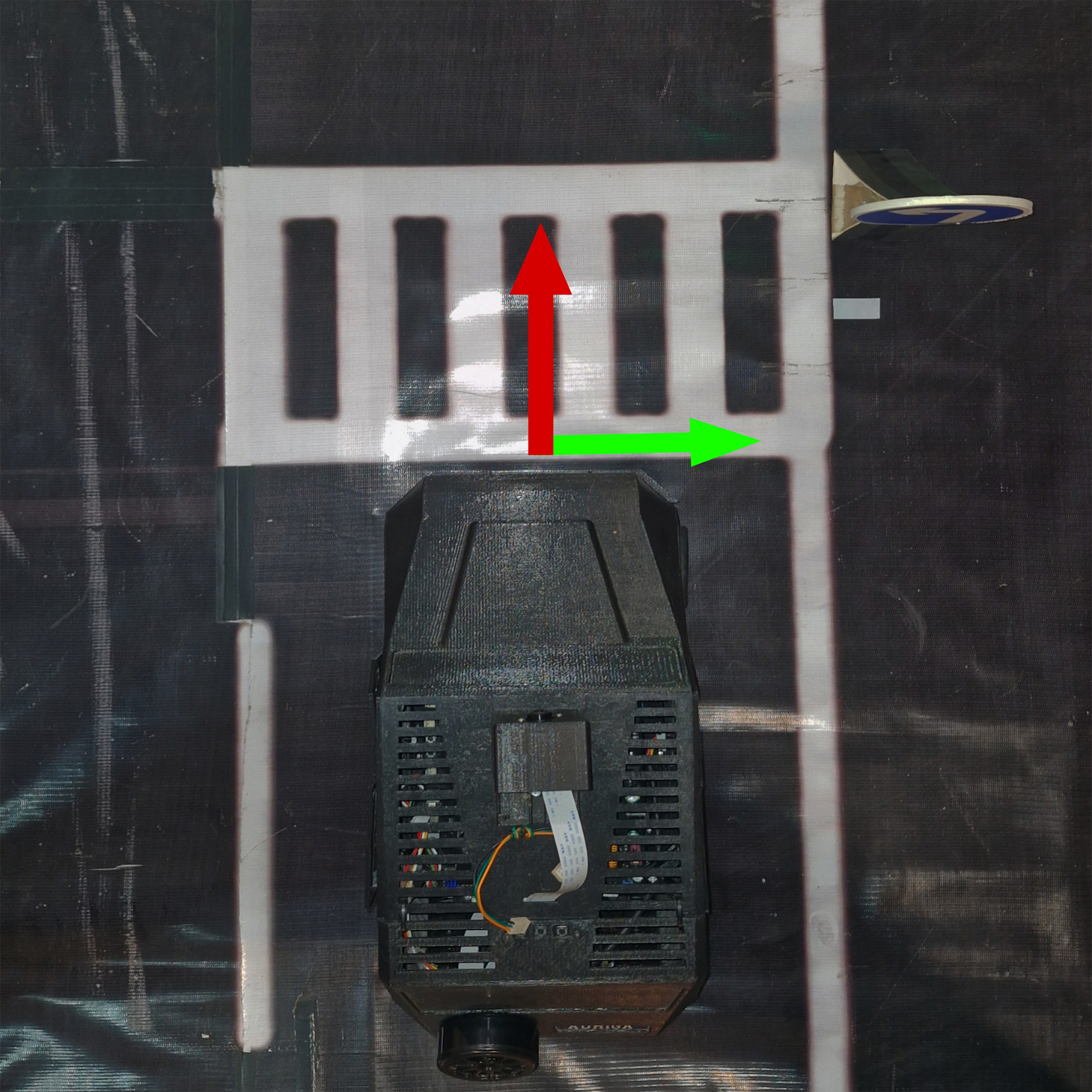}}} %
    \\
    \subfloat[\centering]{{ \includegraphics[width=4cm]{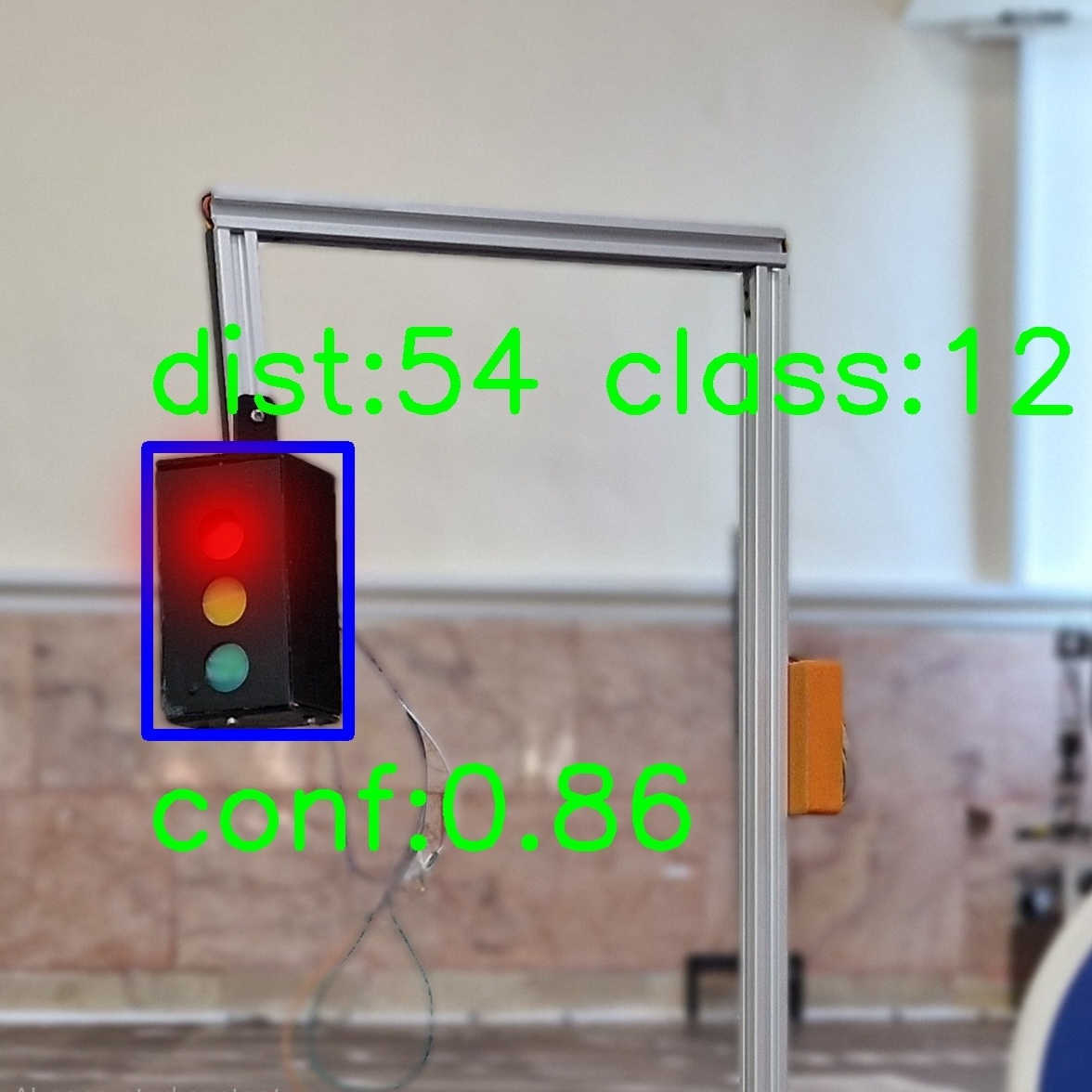}}} %
    \qquad
    \subfloat[\centering]{{ \includegraphics[width=4cm]{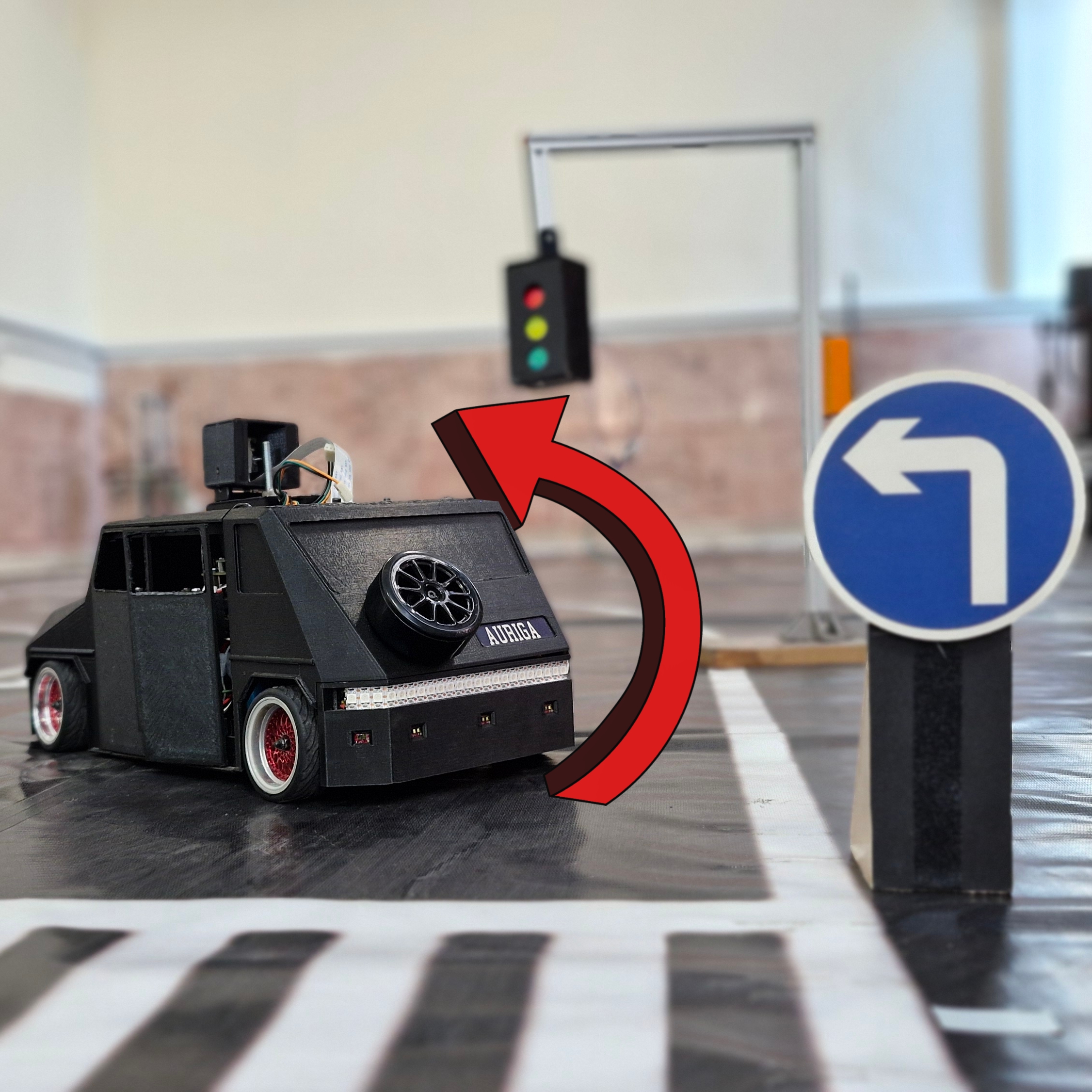}}} %
    \caption{Intersection maneuver sequence: (a) Sign detection; (b) Alignment with the intersection; (c) Traffic light detection; and (d) Executing the turn.} \label{fig16}
\end{figure}

When approaching an intersection, the vehicle responds based on the detected sign, which may indicate a right turn, left turn, or straight movement. The maneuver is executed according to a Finite State Machine (FSM) that encompasses all three possible actions, enabling the vehicle to seamlessly choose the appropriate behavior, as depicted in Figure \ref{fig16}. Before entering the intersection, the vehicle calculates the curvature of the road using gyroscope data to adjust its trajectory. Simultaneously, image processing data is combined with localization information from GPS and other sensors to ensure the vehicle is optimally positioned for safely crossing the intersection with minimal deviation.
Traffic lights detected via image processing take precedence over other maneuvers at intersections, allowing the vehicle to adjust its behavior in real time based on light signals. During this process, all image data undergo statistical filtering using standard deviation to eliminate invalid or noisy information, ensuring accurate decision-making and preventing errors in critical traffic scenarios. This combination of sensor fusion and real-time data filtering allows the vehicle to navigate intersections efficiently while maintaining safety and precision.

\newpage

In addition to handling intersections and parking, the vehicle is programmed to react to other signs such as tunnel warnings, pedestrian crossings, and no-overtaking zones. These reactions include actions like turning on headlights, avoiding overtaking, or activating hazard lights when necessary.

\section{Conclusion}
The Auriga autonomous vehicle exemplifies a well-engineered approach to autonomous navigation, combining advanced hardware components with efficient sensor integration and real-time control algorithms. The vehicle's sophisticated design allows for precise detection of lanes, obstacles, and road signs, ensuring accurate navigation through complex environments. Its control systems, including the Stanley and PID controllers, enhance the vehicle’s responsiveness and stability during operation. The integration of powerful vision and localization technologies positions this AV as a capable and adaptable solution for a wide range of autonomous driving tasks. Future developments will focus on refining the system’s overall efficiency and expanding its ability to handle even more dynamic environments with greater precision.

\newpage

\begin{table}
\centering
\caption{General Characteristics}\label{tab3}
\begin{tabular}{|l|l|}
\hline
Attrbiute &  Value \\
\hline
System Weight & 4.7Kg \\
Overall Length & 39.7cm \\
Overall Width & 24.1cm \\
Overall Height & 20.5cm \\
Max Velocity & 90cm/s \\
Structure Material & PLA \\
Communication & SPI-UART-I2C \\
Voltage connection & 12.6 Vdc \\
\hline
\end{tabular}
\end{table}

%
%
%
%

\bibliographystyle{plain}
\bibliography{citations}

\end{document}